# Fast Quasi-Optimal Power Flow of Flexible DC Traction Power Systems

*Abstract*—This paper proposes a quasi-optimal power flow (OPF) algorithm for flexible DC traction power systems (TPSs). Near-optimal solutions can be solved with high computational efficiency by the proposed quasi-OPF. Unlike conventional OPF utilizing mathematical optimization algorithms, the proposed quasi-OPF adopts analytical mapping from load information to near-optimal solutions, hence considerably accelerating the computation. First, we study the mechanism and physical meaning interpretation of conventional OPF based on a new modeling method and successfully interpret the mechanism of conventional OPF in flexible DC TPSs. Then, the analytical mapping from load information to near-optimal solutions is obtained inspired by the mechanism of conventional OPF, and the quasi-OPF algorithm is designed based on the mapping. Since the mapping is based on simple arithmetic, the quasi-OPF algorithm can solve OPF with much less execution time, achieving subsecond level calculation and a speed-up of 57 times compared to conventional OPF. The effectiveness is verified by mathematical proofs and a case study with Beijing Metro Line 13. It provides an insight into the mechanism and physical meaning of OPF, and is a powerful tool for flexible DC TPSs to analyze the effects of coordinated control, design real-time control strategies, and solve operational problems in planning.

## Nomenclature

*Abbreviations*

| | |
|---|---|
| TPS | Traction power system. |
| TSS | Traction substation. |
| VSC | Voltage sourced converter. |
| OPF | Optimal power flow. |

*Variables*

| | |
|---|---|
| $U_{si}$ | DC voltage of the $i$th TSS. |
| $\boldsymbol{U}_s$ | Vector of TSS DC voltage $U_{si}$. |
| $U_{s\_cm}$ | TSS common mode voltage. |
| $\boldsymbol{U}_{s\_cm}$ | Vector of TSS common mode voltage $U_{s\_cm}$. |
| $\boldsymbol{U}_{s\_cm}{}^*$ | Control order of $\boldsymbol{U}_{s\_cm}$. |
| $U_{si\_dm}$ | The $i$th TSS's differential mode voltage. |
| $\boldsymbol{U}_{s\_dm}$ | Vector of TSS differential mode voltages $U_{si\_dm}$. |
| $\boldsymbol{U}_{s\_dm}{}^*$ | Control order of $\boldsymbol{U}_{s\_dm}$. |
| $U_{ci\_cc}$ | Catenary voltage between the $i$th and the $(i+1)$th TSS in the coordinated control subsystem. |
| $\boldsymbol{U}_{c\_cc}$ | Vector of $U_{ci\_cc}$. |
| $U_{ri\_cc}$ | Rail voltage between the $i$th and the $(i+1)$th TSS in the coordinated control subsystem. |
| $\boldsymbol{U}_{r\_cc}$ | Vector of $U_{ri\_cc}$. |
| $U_{bi\_cc}$ | Branch voltage between the $i$th and the $(i+1)$th TSS in the coordinated control subsystem, i.e., the sum of $U_{ci\_cc}$ and $U_{ri\_cc}$. |
| $\boldsymbol{U}_{b\_cc}$ | Vector of $U_{bi\_cc}$. |
| $\boldsymbol{U}_{b\_cc}{}^*$ | Control order of $\boldsymbol{U}_{b\_cc}$. |
| $I_{si}$ | DC current of the $i$th TSS. |
| $\boldsymbol{I}_s$ | Vector of TSS DC current $I_{si}$. |
| $I_{si\_nd}$ | Natural distribution component of TSS current $I_{si}$, also the current of the $i$th TSS in the natural distribution subsystem. |
| $\boldsymbol{I}_{s\_nd}$ | Vector of $I_{si\_nd}$. |
| $I_{si\_cc}$ | Coordinated control component of TSS current $I_{si}$, also the current of the $i$th TSS in the coordinated control subsystem. |
| $\boldsymbol{I}_{s\_cc}$ | Vector of $I_{si\_cc}$. |
| $\boldsymbol{I}_{s\_cc}{}^*$ | Control order of $\boldsymbol{I}_{s\_cc}$. |
| $I_{ci\_cc}$ | Catenary current between the $i$th and the $(i+1)$th TSS in the coordinated control subsystem. |
| $\boldsymbol{I}_{c\_cc}$ | Vectors of $I_{ci\_cc}$. |
| $\boldsymbol{I}_{c\_cc}{}^*$ | Control order of $\boldsymbol{I}_{c\_cc}$. |
| $I_{ri\_cc}$ | Rail current between the $i$th and the $(i+1)$th TSS in the coordinated control subsystem. |
| $\boldsymbol{I}_{r\_cc}$ | Vector of $I_{ri\_cc}$. |
| $P_{vj}$ | Power of the $j$th rail vehicle. |
| $U_{vj}$ | Voltage of the $j$th rail vehicle. |
| $I_{vj}$ | Current of the $j$th rail vehicle. |
| $P_{cost}$ | Total power obtained from AC utility. |
| $dis$ | Location that varies from 0 to $L$. |

*Constant Coefficients*

| | |
|---|---|
| $N$ | Number of TSSs. |
| $M$ | Number of rail vehicles. |
| $r_{ci}$ | Catenary resistance between the $i$th and the $(i+1)$th TSS in coordinated control subsystem. |
| $r_{ri}$ | Rail resistance between the $i$th and the $(i+1)$th TSS in the coordinated control subsystem. |
| $\boldsymbol{G}$ | Branch conductance matrix in the coordinated control subsystem. |
| $\boldsymbol{R}$ | Branch resistance matrix in the coordinated control subsystem. |
| $I_{si\_lim}$ | TSS $i$'s maximum allowable current. |
| $P_{si\_lim}$ | TSS $i$'s maximum allowable power. |
| $I_{auxi}$ | TSS $i$'s station auxiliary current. |
| $P_{auxi}$ | TSS $i$'s station auxiliary loads power. |
| $L$ | Total length of the metro line. |

## I. Introduction

DC traction power systems (TPSs) are special power systems that energize rail vehicles in urban rail transit systems. The requirements for reliable, low carbon, and economical operation of TPSs are increasingly high [1]. Flexible DC TPSs that introduce voltage sourced converters (VSCs) into DC TPSs are promising for achieving economic and technical merits based on state-of-the-art power electronics and system-level coordinated control. Flexible DC TPSs have high converter-level controllability [2], enable system-level coordinated control strategies to optimize power flow [3], [4], and facilitate the connection to and absorption of renewable energy [5], [6]. Eleven countries with 12 metro lines adopt flexible DC TPSs to achieve better performance [7].

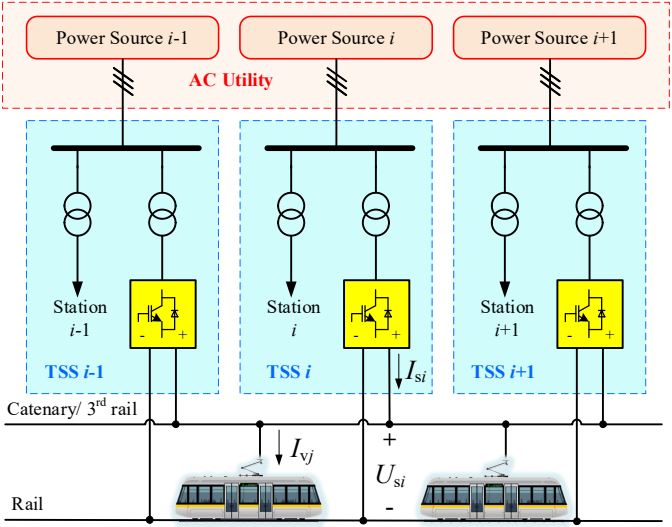

**Fig. 1.** Configuration of flexible DC TPSs.

The optimal power flow (OPF) seeks optimal steady-state setpoints that minimize a certain objective in consideration of various constraints and is widely acknowledged as a powerful tool for analysis, operations, and planning [8], [9]. Since the power flow of flexible DC TPSs can be optimized by system-level coordinated control, OPF is a fundamental tool to analyze the effects of coordinated control and provide an important basis for the design of practical coordinated control strategies.

Flexible DC TPSs are special microgrids, and there are three peculiarities in flexible DC TPSs:

- subsecond change of rail vehicles' power demand. The power demand is very volatile because of sudden acceleration, coasting, or braking.
- time varying rail vehicle positions. Unlike fixed load positions in normal power systems, the loads, i.e., rail vehicles, move along the rails.
- long-chain topology of the DC network. The transmission lines are laid along the rails to power running vehicles. Rails serve as the negative return paths. Hence, transmission lines and rail form a long-chain topology.

The former two peculiarities both highlight the importance of OPF computational efficiency. First, the rapid change of rail vehicles' power demand necessitates a computationally efficient algorithm in real-time coordinated control. OPF must be conducted at the subsecond level for proper application in coordinated control. Otherwise, the dated OPF results may deteriorate the control performance. Second, millions of operating scenarios need to be calculated by OPF in power flow analysis and planning in consideration of all the possible rail vehicle positions, contingency conditions, and departure intervals. Assuming OPF can be solved in 1 s, approximately 100 hours are needed to complete the thorough calculation, which is undesirable. In conclusion, solving the OPF problem with faster speed is significant in flexible DC TPSs.

The OPF of flexible DC TPSs is studied in [10] to achieve better energy management. Similarly, the OPF of DC TPSs considering the parallel running of diode rectifiers and VSCs is studied in [11]-[13]. These studies effectively realize the reliable and economical operation of DC TPSs but do not focus on the computational efficiency problem, and information about the execution time has not been reported. In addition, VSCs can achieve reactive power compensation in the AC network connected to DC TPSs, and the relevant AC OPF is well addressed in [14], [15]. However, our study focuses on the OPF of DC networks.

Various methods have been proposed to speed up the OPF calculation in power systems. Parallel computing is utilized to accelerate the computing speed of various algorithms [16], [17]. A parametric distribution optimal power flow method is proposed in [18] to improve computational efficiency, and the optimal setpoints can be calculated by analytical functions of the volatile renewable output. Machine learning based data-driven methods are a promising approach to increase computational efficiency [19]-[21]; however, the TPS topology is time varying due to the rapid change of rail vehicle positions, and solving OPF at the subsecond level with time varying topologies is still in the initial stage of exploration [22]. Sparsity is exploited to accelerate the computing speed in [23]. Linear approximation methods are adopted in [24] and [25] to simplify the nonlinear OPF problem. Reference [26] proposes an interior point Powerball algorithm to improve the search directions and to accelerate the OPF solution process. Especially aiming at real-time control problems, computationally efficient OPF algorithms have been developed in consideration of real-time measurement and communication [27]-[30].

A new methodology is adopted in our study to achieve high computational efficiency in flexible DC TPSs. We study the mechanism and physical meaning interpretation of OPF based on large quantities of observations on the results of OPF and successfully interpret the mechanism of OPF considering the special long-chain topology in flexible DC TPSs: only neighboring traction substations (TSSs) provide coordinated support to achieve optimal operation, and the coordinated support is assigned to neighboring TSSs reverse proportionally to the distances. Then, an analytical mapping from load information to near-OPF solutions is designed based on the OPF mechanism. The mapping only needs simple arithmetic, hence contributing to high computational efficiency. Based on the analytical mapping, a quasi-OPF algorithm is proposed. Quasi-OPF is an algorithm that solves near-optimal solutions for OPF without using mathematical optimization algorithms. Quasi-OPF shares many key similarities with conventional OPF:

- the solutions are almost the same as or very close to conventional OPF's,
- the decision variables, objective, and considered constraints are the same as conventional OPF's.

However, the proposed quasi-OPF adopts explicit mapping from load information to OPF solutions rather than using regular mathematical optimization algorithms. Hence, the proposed algorithm considerably increases the computational efficiency.

The twofold contributions can be summarized as follows:

First, we reveal the mechanism of OPF based on a new modeling method. The modeling method is based on the superposition principle, which decomposes an original system into the superposition of a coordinated control subsystem and a natural distribution subsystem. The power flow of the coordinated control subsystem only depends on the effect of coordinated control, whereas the power flow of the natural distributed subsystem only depends on the natural distribution of the power demand of rail vehicles. The effect of coordinated control can be revealed from the complex power flow by this system modeling method, which helps to reveal the mechanism of OPF.

Second, we propose a quasi-OPF algorithm based on the mechanism analysis of OPF. This algorithm simulates OPF with high computational efficiency, which achieves a speed-up of 57 times. The effectiveness is verified by mathematical proofs and a case study with Beijing Metro Line 13.

This study is beneficial to real-time control and planning of flexible DC TPSs. Since the proposed model can generate steady-state near-optimal control orders rapidly, a practical real-time control strategy can be designed based on the proposed model. However, a real-time control strategy needs to handle many other problems other than steady-state optimization, such as communication network deployment and dynamic process consideration. Our proposed model cannot be used directly in real-time control but provides an important basis for it. Furthermore, many planning frameworks utilize OPF in lower-level optimization to solve the operational problem [31]-[34]. In this case, our proposed model can be utilized to solve the operational optimization problem in planning.

This paper is organized as follows. In Section II, the flexible DC TPSs' basic configuration and OPF are introduced as preliminaries. In Section III, the system modeling based on the superposition principle is elaborated, and the internal mechanism of OPF is analyzed. In Section IV, the fast quasi-OPF algorithm is proposed. In Section V and Section VI, a case study and conclusions are provided, respectively. The mathematical proofs are elaborated in the appendix.

## II. OPF OF FLEXIBLE DC TPSS

### A. Configuration of Flexible DC TPSs

The basic configuration of the flexible DC TPS is shown in Fig. 1. VSCs are adopted to supply power to rail vehicles. In each TSS, a bus is energized by a power source from AC utility and supplies power to transformers. There are two kinds of transformers in TSSs: one supplies power to VSCs, and the other energizes station auxiliary loads, such as station lighting, elevators, ventilation, *etc*. The station auxiliary loads are usually considered constant loads. VSCs, catenaries, rail vehicles, and rails form loops to transmit electric power.

### B. Control Objective

To minimize operating costs, the objective is defined by:

$$P_{\text{cost}}(\bm{U}_\text{s}) = \sum_{i=1}^{N} \max(P_{si} + P_{\text{aux}i}, 0) \quad (1)$$

where $P_{\text{cost}}$ is the total power obtained from AC utility, which is also the power charged by utility, $N$ is the number of TSSs, $P_{si}$ is the DC VSC power of the $i$th TSS, and $P_{\text{aux}i}$ is the station auxiliary load power of the $i$th TSS. $\bm{U}_\text{s}$ is defined as the vector of TSS DC voltage:

$$\bm{U}_\text{s} = [U_{s1} \cdots U_{si} \cdots U_{sN}]^\text{T} \quad (2)$$

where $U_{si}$ is the DC voltage of the $i$th TSS. Since the voltage of VSCs can be controlled flexibly, the control variable is $\bm{U}_\text{s}$. The reason why the $\max(x,0)$ function is applied in (1) is explained as follows.

Modern rail vehicles can realize bidirectional flow of electric power: they consume electric power when they are accelerating or cruising, whereas they regenerate electric power when they convert mechanical energy to electric power by regenerative braking. Hence, reducing energy consumption is equivalent to two goals: to recuperate more rail vehicles' regenerative power and to lower power loss. Lowering power loss is important but less significant than recuperating regenerative power because the power loss is considerably less than the regenerative power.

Rail vehicles' regenerative power may flow to three kinds of terminals: other rail vehicles, station auxiliary loads, or AC utility. It is a priority for TPSs to absorb regenerative power by other rail vehicles and station auxiliary loads because power transmitted back to AC utility is discouraged by legislations and rules in practice [13]: it results in difficulty for AC utility to absorb and manage, it increases energy consumption, and the power quality may be undesirable. For countries such as China, the transit property cannot obtain revenue and may even have to pay fines to transmit power back to AC utility. Therefore, only the energy obtained from AC utility is charged in most cases, and the $\max(x,0)$ function in (1) represents this practical charge rule.

### C. Operational Constraints

To ensure reliable operation, there are two main constraints: the peak power of VSCs and the range of DC voltage.

The peak power of VSCs is limited to ensure VSC safety. In normal cases, the constraint of VSC peak power is more rigid than the constraint of power line peak power.

The DC voltage is limited to ensure reliable operation of rail vehicles. In TPSs whose rated voltage is 750 V, the voltage range is usually 500 V-900 V. When braking rail vehicles are regenerating power back to catenaries, the maximum allowable voltage for rail vehicles is changed from 900 V to 950 V.

### D. OPF

The objective is (1). The constraints of OPF include equality constraints, i.e., power balance equations, and inequality constraints about operational constraints: VSC power limitation and DC voltage limitation. The OPF problem is formulated as:

$$\min_{\bm{U}_\text{s}} P_{\text{cost}}(\bm{U}_\text{s}) = \sum_{i=1}^{N} \max(P_{si} + P_{\text{aux}i}, 0)$$

$$\text{s.t.} \quad P_{vj} = U_{vj} \sum_{i=1}^{N+M} G_{nji} U_i, \; j=1,2,\cdots,M \quad (3)$$

$$\underline{P_{si}} \leq P_{si} \leq \overline{P_{si}}, \; i=1,2,\cdots,N$$

$$\underline{U_i} \leq U_i \leq \overline{U_i}, \; i=1,2,\cdots,N+M$$

where $P_{vj}$ and $U_{vj}$ are the power and voltage of the $j$th rail vehicle, respectively; $M$ is the number of rail vehicles, $G_{nji}$ is the $(j,i)$th element in the node conductance matrix, $U_i$ is the voltage of the $i$th bus; $\overline{U_i}$ and $\overline{P_{si}}$ are the upper bounds of voltage of bus $i$ and power of TSS $i$, respectively; $\underline{U_i}$ and $\underline{P_{si}}$ are the lower bounds of voltage of bus $i$ and power of TSS $i$, respectively.

The OPF problem can be solved by the primal-dual interior point method in MATLAB. Note that the max($x$,0) function in (1) is nonlinear, which complicates the calculation of OPF. Approximately 10 times can be sped up without the max($x$,0) function in the objective function.

### III. SYSTEM MODELING AND ANALYSIS

*A. System Modeling*

In system modeling, an original system is shown in Fig. 2(a). We assume rail vehicles as current sources and TSSs as voltage sources. For the $i$th TSS, its DC voltage and current are $U_{si}$ and $I_{si}$, respectively. For the $j$th rail vehicle, its DC current is $I_{vj}$. Note that the parallel up-tracks and down-tracks are omitted in Fig. 2 for simplicity.

Since the time scale of VSC control is much shorter than the time scale of power flow change, the detailed internal operation and control of VSCs are not considered in the OPF model. In our study, VSCs are viewed as ideal voltage sources.

In normal OPF or power flow modeling, the loads are usually modeled as constant power nodes, and the power value is known by the input. However, constant power nodes introduce nonlinearity in the system modeling, which is difficult to do theoretical analysis. Therefore, based on the substitution theorem in circuit theory, we model the loads, namely, the rail vehicles, as current sources, and the current value will be determined later to ensure that the load power is equal to the input value. In conclusion, the current sources of rail vehicles reflect the load power.

We use $\boldsymbol{U}_s$ and $\boldsymbol{I}_s$ to denote the vector of TSS voltage and current, respectively. $\boldsymbol{U}_s$ can be decomposed into two components, as shown in Fig. 2(b), which supposes:

$$\boldsymbol{U}_s = \boldsymbol{U}_{s\_cm} + \boldsymbol{U}_{s\_dm} \quad (4)$$

where $\boldsymbol{U}_{s\_cm}$ and $\boldsymbol{U}_{s\_dm}$ are the vectors of TSS common mode and differential mode voltages, respectively. Apparently, the original system in Fig. 2(a) is equivalent to the system in Fig. 2(b). $U_{si\_dm}$ indicates the $i$th TSS's differential mode voltage; $U_{s\_cm}$ indicates the TSS common mode voltage. Assume there are $N$ TSSs in the whole line. For simplicity, we assume that the voltage of the $N$th TSS is the common mode voltage $U_{s\_cm}$.

Based on the superposition principle in circuit theory, the system in Fig. 2(b) is equivalent to the superposition of the natural distribution subsystem in Fig. 2(c) and the coordinated control subsystem in Fig. 2(d). The natural distribution subsystem only reserves the voltage sources of TSS common mode voltage and the current sources of rail vehicles. The coordinated control subsystem only reserves the voltage sources of the TSS differential mode voltage. Note that blue variables are related to the natural distribution subsystem, whereas the red variables are related to the coordinated control subsystem in Fig. 2.

The "natural distribution" means the power flow exclusively excited by the current sources of rail vehicles. There are two kinds of excitation sources in the DC TPSs, namely, the voltage sources of TSSs and the current sources of rail vehicles. The voltages of TSSs are control variables that reflect the control effect, whereas the currents of rail vehicles reflect the power of loads. The power flow exclusively excited by the current sources of rail vehicles is not influenced by the control effect but represents the natural power flow distribution of loads. In the natural distribution subsystem, as shown in Fig. 2(c), voltage sources of TSS common mode voltage $U_{s\_cm}$ cannot excite power flow because all TSSs have the same voltage $U_{s\_cm}$; if there is no rail vehicle, there will be no current through any branches. Only the current sources of rail vehicles act as excitation sources. Therefore, the power flow of the natural distributed subsystem represents the natural power flow distribution of loads.

In the coordinated control subsystem shown in Fig. 2(d), the power flow is exclusively excited by $\boldsymbol{U}_{s\_dm}$. Since the control variables are VSC voltages $\boldsymbol{U}_s$ and $\boldsymbol{U}_{s\_cm}$ cannot excite power flow, the coordinated control can only be excited by $\boldsymbol{U}_{s\_dm}$. Therefore, the power flow of the coordinated control subsystem is dependent on coordinated control, reflecting the coordinated control effect.

In the natural distributed subsystem, the $i$th TSS's current is $I_{si\_nd}$, which is the natural distribution component of TSS current $I_{si}$. In the coordinated control subsystem, the $i$th TSS's current is $I_{si\_cc}$, which is the coordinated control component of TSS current $I_{si}$. Based on the superposition principle, $\boldsymbol{I}_s$ can also be decomposed into two components:

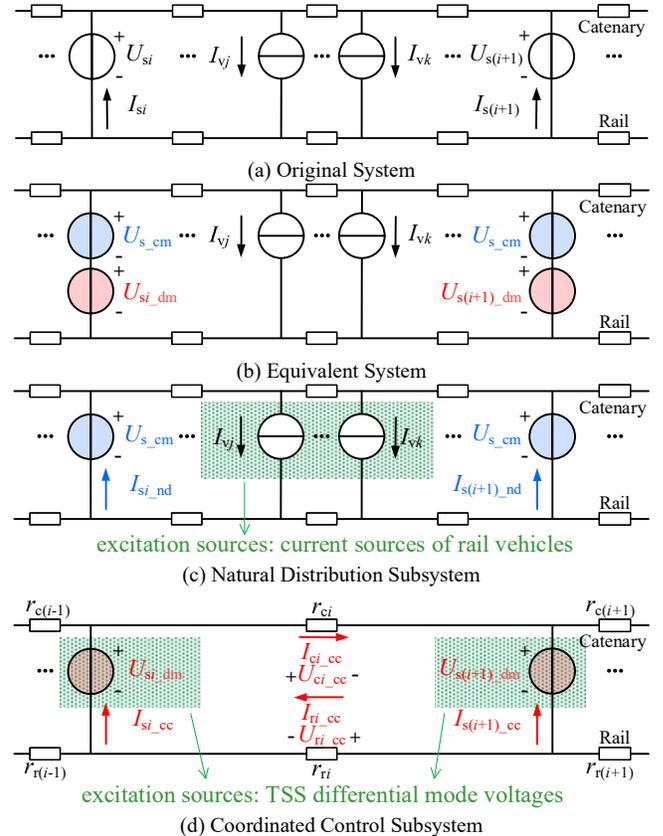

**Fig. 2.** System modeling based on the superposition principle.

$$I_s = I_{s\_nd} + I_{s\_cc} \quad (5)$$

where $I_{s\_nd}$ and $I_{s\_cc}$ are the vectors of $I_{si\_nd}$ and $I_{si\_cc}$, respectively.

In the coordinated control subsystem shown in Fig. 2(d), considering the branch between the $i$th TSS and the $(i+1)$th TSS, $r_{ci}$ and $r_{ri}$ denote the catenary and rail resistance, respectively; $I_{ci\_cc}$ and $U_{ci\_cc}$ denote the catenary current and voltage, respectively; and $I_{ri\_cc}$ and $U_{ri\_cc}$ denote the rail current and voltage, respectively. Since this subsystem is a two-port network, based on Kirchhoff's current law:

$$I_{c\_cc} = I_{r\_cc} \quad (6)$$

where $I_{c\_cc}$ and $I_{r\_cc}$ are the vectors of $I_{ci\_cc}$ and $I_{ri\_cc}$, respectively.

Define branch voltage $U_{bi\_cc}$ to simplify the analysis:

$$U_{b\_cc} = U_{c\_cc} + U_{r\_cc} \quad (7)$$

where $U_{b\_cc}$, $U_{c\_cc}$, and $U_{r\_cc}$ denote the vectors of $U_{bi\_cc}$, $U_{ci\_cc}$, and $U_{ri\_cc}$, respectively. Based on Kirchhoff's voltage law:

$$U_{b\_cc} = \begin{bmatrix} 1 & -1 & 0 & \cdots & 0 & 0 \\ 0 & 1 & -1 & \cdots & 0 & 0 \\ 0 & 0 & 1 & \cdots & 0 & 0 \\ \vdots & \vdots & \vdots & \ddots & \vdots & \vdots \\ 0 & 0 & 0 & \cdots & -1 & 0 \\ 0 & 0 & 0 & \cdots & 1 & -1 \end{bmatrix} U_{s\_dm} \quad (8)$$

Define conductance matrix $G$:

$$G = diag((r_{c1}+r_{r1})^{-1}, (r_{c2}+r_{r2})^{-1}, \ldots, (r_{c(N-1)}+r_{r(N-1)})^{-1}) \quad (9)$$

Based on Ohm's law and Kirchhoff's current law:

$$I_{c\_cc} = G U_{b\_cc} \quad (10)$$

$$I_{s\_cc} = \begin{bmatrix} 1 & 0 & 0 & \cdots & 0 & 0 \\ -1 & 1 & 0 & \cdots & 0 & 0 \\ 0 & -1 & 1 & \cdots & 0 & 0 \\ \vdots & \vdots & \vdots & \ddots & \vdots & \vdots \\ 0 & 0 & 0 & \cdots & -1 & 1 \\ 0 & 0 & 0 & \cdots & 0 & -1 \end{bmatrix} I_{c\_cc} \quad (11)$$

*B. OPF Mechanism Analysis*

The mechanism analysis of OPF can be illustrated by a typical OPF result, as shown in Fig. 3. The parameters and descriptions of the case are given in the appendix. Based on the system modeling, the optimal solution can be interpreted from three aspects:

1) **Limiting TSS Peak Power**

$I_{s2\_nd}$ is large, which means that the power demand of rail vehicles near TSS 2 is large. TSS 2 would violate the peak power constraint without coordinated control.

The optimal TSS voltages form a hollow in the area of TSS 1-3 to limit the peak power of TSS 2. Since the voltages of TSS 1 and 3 are higher than TSS 2, TSS 1 and 3 output more power to energize rail vehicles near TSS 2, and hence TSS 2 can lower its power output.

2) **Recuperating Regenerative Power**

$I_{s20\_nd}$ is negative and large, which means that there is large regenerative power near TSS 2. TSS 2 cannot recuperate all the naturally distributed regenerative power without coordinated control.

The optimal TSS voltages form a peak in the area of TSS 19-21 to recuperate the regenerative power near TSS 20. Since the voltages of TSS 19 and 21 are lower than TSS 20, the braking rail vehicles can transmit their regenerative power to rail vehicles near TSS 19 and 21, and hence TSS 20 can avoid transmitting power back to AC power sources.

3) **Lowering Power Loss**

As observed in Fig. 3, the coordinated support only comes from neighboring TSSs. In addition, the coordinated support is assigned to neighboring TSSs inversely proportional to the branch resistances between the target TSS and each TSS, i.e.,

$$I_{s1\_cc} : I_{s3\_cc} \approx (r_{c2} + r_{r2}) : (r_{c1} + r_{r1}) \quad (12)$$

$$I_{s19\_cc} : I_{s21\_cc} \approx (r_{c21} + r_{r21}) : (r_{c19} + r_{r19}) \quad (13)$$

This pattern can be referred to as a proximity principle.

The proximity principle can be understood from the aspects of power loss and voltage drop. If the proximity principle does not hold true, then the further the supportive TSS is, the more coordinated support it provides; then $I_{s\_cc}$ would flow through a larger branch resistance, resulting in larger power loss and larger voltage fluctuation.

In conclusion, the coordinated control component of TSS current $I_{s\_cc}$ can be decomposed from TSS current $I_s$ based on the proposed system modeling. $I_{s\_cc}$ reveals the essence of coordinated support. The optimal $I_{s\_cc}$ covers the demands to satisfy the peak power constraint and to minimize the objective. Control variables in $U_s$ are the means to generate optimal $I_{s\_cc}$.

## IV. QUASI-OPF ALGORITHM

The quasi-OPF algorithm is inspired by the mechanism of OPF. It essentially is the simulation of $I_{s\_cc}$ utilization in OPF.

*A. Overview of the Algorithm*

The flow chart of the algorithm is shown in Fig. 4. For clarity, for variables as reference orders, the corner mark $^*$ is added as a suffix. There are four steps in the algorithm, which will be elaborated as follows:

The input data include traction calculation results, i.e., speed/location/power of rail vehicles versus time, physical parameters, and operational requirements and constraints.

The first step is power flow calculation, which can be conducted by mature algorithms [35], [36].

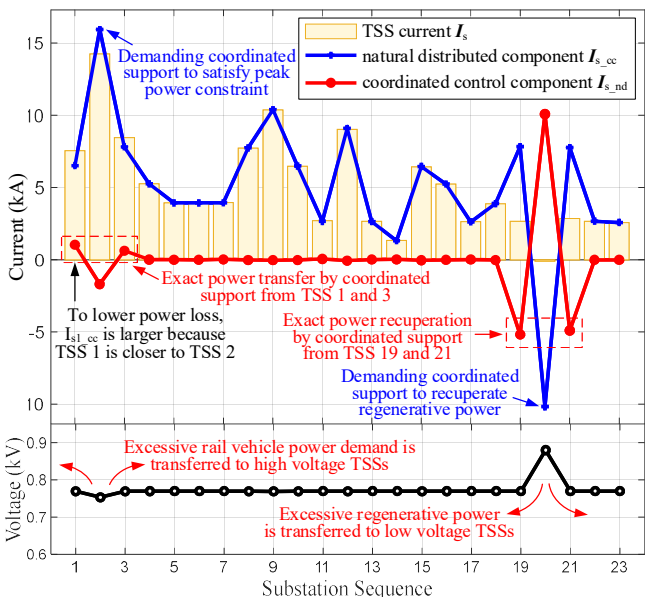

**Fig. 3.** OPF result and mechanism analysis at a typical instant.

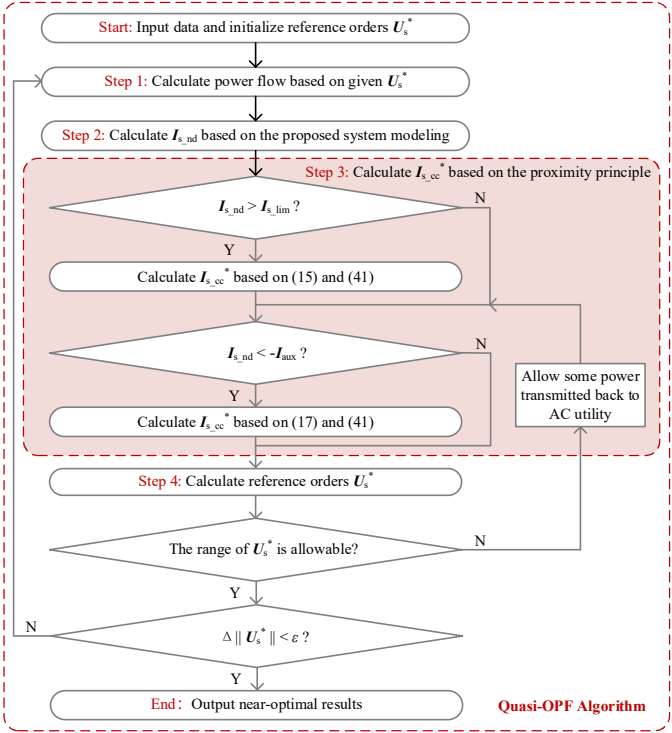

**Fig. 4.** Flow chart of the quasi-OPF algorithm.

The second step is based on the system modeling. The TSS current $I_s$ and voltage $U_s$ are calculated in step 1. Assuming the voltage of the $N$th TSS is the common mode voltage $U_{s\_cm}$, we can calculate $U_{s\_dm}$ by (4). Then, $I_{s\_cc}$ can be calculated using (8), (10), and (11), and $I_{s\_nc}$ can be calculated by (5).

### B. Step 3: Calculation of Current Reference $I_{s\_cc}^*$

Step 3 is the most significant part of this algorithm. By simulating the mechanism of OPF, the ideal reference order $I_{s\_cc}^*$ can be calculated based on $I_{s\_nc}$ from step 2. Similar to the mechanism analysis, $I_{s\_cc}^*$ is solved considering three aspects:

1) **Limiting TSS Peak Power**

$I_{si\_lim}$ is used to denote TSS $i$'s maximum allowable current, which can be calculated by:

$$I_{si\_lim} = P_{si\_lim} / U_{si} \tag{14}$$

where $P_{si\_lim}$ is TSS $i$'s maximum allowable power.

Assuming $I_{si\_nc}$ is larger than $I_{si\_lim}$, supportive TSSs should increase their DC voltage and transmit more current to rail vehicles near TSS $i$. Then, $I_{si\_cc}^*$ is determined by

$$I_{si\_cc}^* = I_{si\_lim} - I_{si\_nd} \tag{15}$$

which means that the excessive power demand of rail vehicles near TSS $i$ is energized by supportive TSSs, and the output power of TSS $i$ is limited exactly at the maximum allowable value.

2) **Recuperating Regenerative Power**

$I_{auxi}$ is used to denote TSS $i$'s station auxiliary current, which can be calculated by:

$$I_{auxi} = P_{auxi} / U_{si} \tag{16}$$

where $P_{auxi}$ is TSS $i$'s station auxiliary loads.

Assuming $I_{si\_nc}$ is lower than $I_{auxi}$, TSS $i$ should increase its DC voltage to ensure that braking rail vehicles can transmit their regenerative power to rail vehicles near supportive TSSs. Then, $I_{si\_cc}^*$ is determined by

$$I_{si\_cc}^* = -I_{auxi} - I_{si\_nd} \tag{17}$$

which means that the excessive regenerative power of rail vehicles near TSS $i$ is transferred to other rail vehicles and that there is exactly no power transmitted back to AC utility from TSS $i$.

It is worth mentioning that the range of $U_s^*$ may violate the constraint. Some regenerative power can be allowed to transmit to AC utility to satisfy the DC voltage constraint.

3) **Lowering Power Loss**

$I_{s\_cc}^*$ of supportive TSSs should be optimized to lower power loss and is calculated based on the proximity principle observed in the mechanism analysis of OPF. The proximity principle can be interpreted as follows: only neighboring TSSs provide coordinated support; the closer the supportive TSS is, the more coordinated support it provides.

The effectiveness of the proximity principle can be proved by mathematical derivation, which is elaborated as follows:

*Proposition 1*: In the sense of probability statistics, optimizing power loss in the coordinated control subsystem is equivalent to optimizing power loss in the original system.

The proof is shown in the appendix. Note that the assumption made in the proof of Proposition 1 leads to some difference between the quasi-OPF and the actual OPF results, which is further illustrated in the case study.

*Proposition 2*: To minimize power loss, if TSS ($m$-1) and ($m$+1) do not need coordinated support, then $I_{sm\_cc} = 0$.

In other words, $I_{si\_cc}$ is nonzero when and only when its neighboring TSSs need coordinated support.

The proof is shown in the appendix. In the process of proof, a more powerful corollary can be derived:

*Corollary 1*: If TSS ($m$-1) and ($m$+1) do not need coordinated support, then branch currents $I_{c(m-1)\_cc}$ and $I_{cm\_cc}$ are 0, which means that the left side of the TSS ($m$-1) is isolated from the right side of the TSS ($m$+1).

The proof is shown in the appendix. Corollary 1 means that the TSSs on the left side of the TSS ($m$-1) or the TSSs on the right side of the TSS ($m$+1) are autonomous. The two parts do not affect each other at all. Then, the last proposition can be derived based on corollary 1:

*Proposition 3*: To minimize power loss, if TSS $m$ needs coordinated support, whereas TSS $p$ and $q$ are supportive TSSs, then the coordinated support is assigned to TSS $p$ and $q$ reverse proportionally to the branch resistances between TSS $m$ and each TSS.

The proof is shown in the appendix.

In conclusion, based on Propositions 1, 2, and 3, we prove that adopting the proximity principle to calculate $I_{s\_cc}^*$ of supportive TSSs can optimize power loss in the sense of probability statistics.

### C. Step 4: Calculation of Voltage Reference $U_s^*$

Step 4 is basically the inverse process of step 2. $I_{c\_cc}^*$ can be calculated based on (25) and $I_{s\_cc}^*$, which are known in step 3. $U_{b\_cc}^*$ can be calculated by:

$$U_{b\_cc}^* = RI_{c\_cc}^* \tag{18}$$

Then, $U_{s\_dm}^*$ can be calculated:

$$U_{s\_dm}{}^* = \begin{bmatrix} 1 & 1 & \cdots & 1 & 1 \\ 0 & 1 & \cdots & 1 & 1 \\ \vdots & \vdots & \ddots & \vdots & \vdots \\ 0 & 0 & \cdots & 1 & 1 \\ 0 & 0 & \cdots & 0 & 1 \\ 0 & 0 & \cdots & 0 & 0 \end{bmatrix} U_{b\_cc}{}^* \qquad (19)$$

$U_{s\_cm}{}^*$ can be determined by various methods. In this paper, $U_{s\_cm}{}^*$ is determined by rendering the highest TSS voltage reaching the maximum allowable voltage to make the TSS voltage as high as possible to lower the power loss. Then, the final reference orders for VSCs can be calculated:

$$U_s{}^* = U_{s\_cm}{}^* + U_{s\_dm}{}^* \qquad (20)$$

## V. CASE STUDY

### A. Comparisons in the Whole Operation Cycle

The comparisons are conducted to verify the near-optimal performance of the proposed quasi-OPF algorithm. We run the simulation for 5439 seconds that is the whole running cycle of rail vehicles in Beijing Metro Line 13. The comparisons are recorded in TABLE I, where the energy consumption represents the objective in (1), and the regenerative power recuperation rate is defined as the ratio between the regenerative power recuperated by other rail vehicles and the total regenerative power.

The energy consumption optimization performance and voltage range are very close, and all operation constraints are satisfied, as shown in TABLE I, indicating that the proposed algorithm simulates the OPF successfully.

### B. Comparisons in Typical Instants

The comparisons of calculation results at four typical instants are elaborated in detail to further verify the near-optimization of the proposed algorithm solutions.

The simulation of the OPF mechanism for limiting TSS peak power and lowering power loss can be shown by the comparison of results at typical instant 1. Typical instant 1 is the 3518th second in the whole running cycle. The calculation results are shown in Figs. 5 and 6, respectively. The calculation results are nearly identical.

TABLE I
PERFORMANCE COMPARISONS IN THE WHOLE OPERATION TIME

| Performance Index | QUASI-OPF | OPF |
|---|---|---|
| Energy consumption (10^4 kWh) | 6.343 | 6.328 |
| Regenerative power recuperation rate | 91.76% | 91.80% |
| Voltage range of TSSs (kV) | 0.65-0.88 | 0.65-0.88 |
| Voltage range of rail vehicles (kV) | 0.58-0.94 | 0.58-0.94 |
| Maximum power of VSCs (MW) | 11.0 | 11.0 |

As shown in Fig. 5 or Fig. 6, $I_{s2\_nc}$ and $I_{s18\_nc}$ are so large that VSCs in TSSs 2 and 18 would violate the peak power constraint without coordinated control. The power of the VSCs in TSSs 2 and 18 is limited at the maximum allowable value by the coordinated support from neighboring TSSs.

In addition, the effectiveness of the proximity principle is verified by the $I_{s\_cc}{}^*$ of OPF, as shown in Fig. 6. The power loss of the proposed quasi-OPF is 1.404 MW, whereas the power loss of OPF is 1.397 MW, indicating the small difference in the optimization performance between the two algorithms.

The simulation of the OPF mechanism concerning recuperating regenerative power can be shown by the comparison of results at typical instant 2. Typical instant 2 is the 2508th second in the whole running cycle. The calculation results of the proposed quasi-OPF and OPF are shown in Figs. 7 and 8, respectively. The calculation results are very close.

As shown in Fig. 7 or Fig. 8, $I_{s3\_nc}$, $I_{s4\_nc}$, $I_{s17\_nc}$, $I_{s18\_nc}$, and $I_{s22\_nc}$ are negative and lower than the corresponding $I_{auxi}$, indicating that TSS 3, 4, 17, 18, and 22 would not recuperate all the regenerative power without coordinated control. The power of the VSCs in TSSs 3, 4, 17, 18, and 22 can exactly energize the station auxiliary loads by the coordinated support from neighboring TSSs, which means all the regenerative power is successfully recuperated and there is no power transmitted back to the AC utility.

In addition, there are small differences regarding $I_{s\_cc}{}^*$ and $U_s{}^*$ in the area from TSS 5 to 16, as shown in Fig. 8, because OPF can better optimize power loss in local areas. The power loss of the proposed quasi-OPF is 1.923 MW, whereas the power loss of OPF is 1.858 MW, indicating the small difference in the power loss optimization performance between the two algorithms. As stated above, adopting the proximity principle to calculate $I_{s\_cc}{}^*$ only optimizes power loss in the sense of probability statistics. OPF can better lower the power loss in a given operation scenario compared to the proposed algorithm.

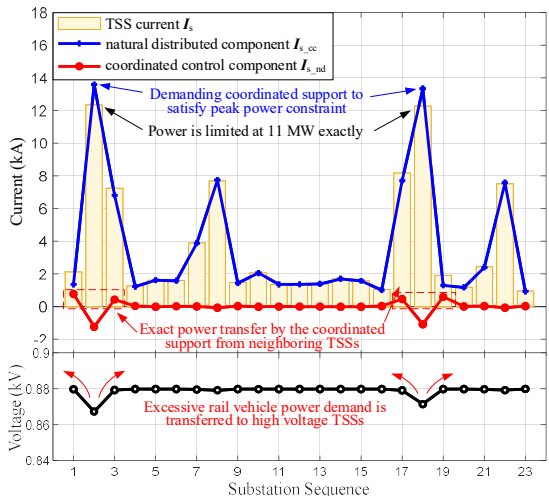

**Fig. 5.** The proposed quasi-OPF result at typical instant 1.

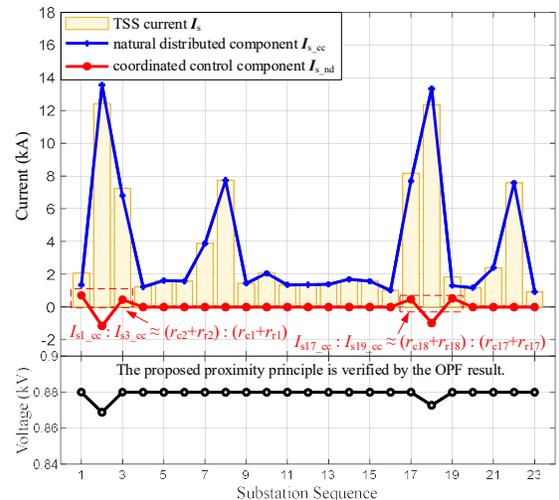

**Fig. 6.** The OPF result at typical instant 1.

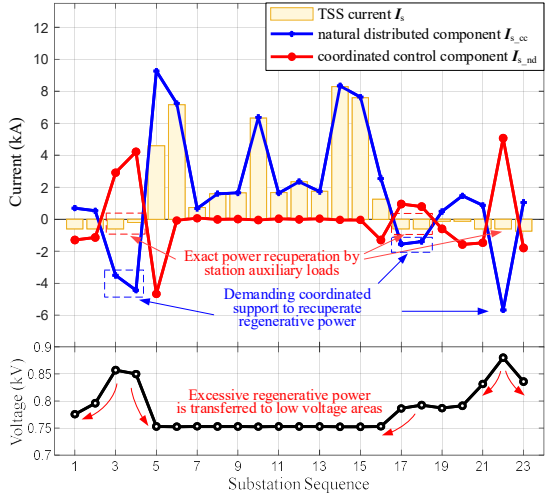

**Fig. 7.** The proposed quasi-OPF result at typical instant 2.

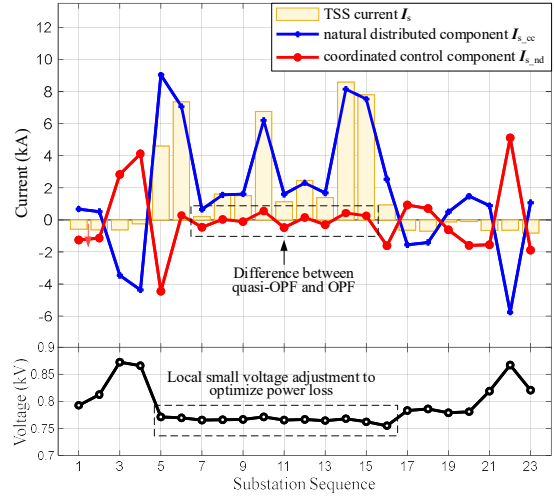

**Fig. 8.** The OPF result at typical instant 2.

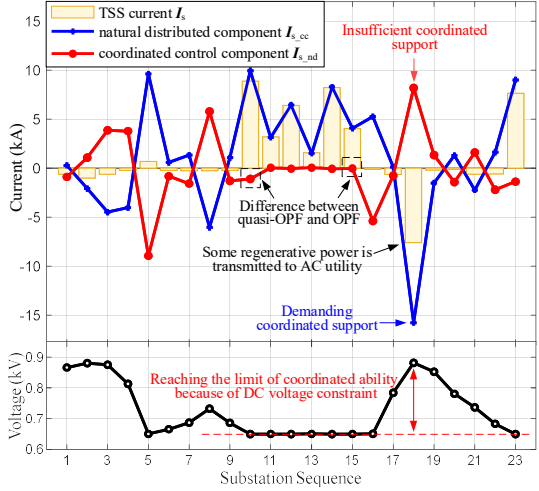

**Fig. 9.** The proposed quasi-OPF result at typical instant 3.

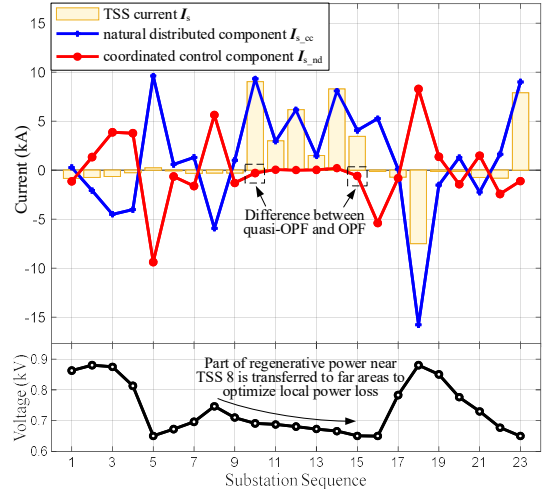

**Fig. 10.** The OPF result at typical instant 3.

A special case in which not all the regenerative power is recuperated is shown by the comparison of results at typical instant 3. Typical instant 3 is the 917th second in the whole running cycle. The calculation results of the proposed quasi-OPF and OPF are shown in Figs. 9 and 10, respectively. The calculation results are very close.

As shown in Fig. 9 or Fig. 10, $I_{s18\_nc}$ is negative and very large. TSS 18 cannot obtain more coordinated support because there is no extra voltage fluctuation range. Hence, some regenerative power is transmitted back to the AC utility from TSS 18. Under this condition, extra calculation is needed to determine how much regenerative power is allowed to be transmitted to the AC utility and calculate voltage references again.

In addition, there are small differences regarding $I_{s\_cc}^*$ of TSS 10 and 15 and $U_s^*$ in the area from TSS 9 to 15, as shown in Fig. 10, because OPF can better optimize power loss in local areas. The power loss of the proposed quasi-OPF is 6.136 MW, whereas the power loss of OPF is 6.092 MW, indicating the small difference in the power loss optimization performance between the two algorithms.

*C. Computational Efficiency Comparisons*

The computational efficiency comparisons are listed in TABLE II. The proposed quasi-OPF algorithm takes 1/57 of the time of OPF. All computations are carried out using MATLAB on a computer with an Intel(R) Core(TM) i7-10700 CPU and 64 GB RAM.

TABLE II
COMPUTATIONAL EFFICIENCY COMPARISONS

| EXECUTION TIME (S) | QUASI-OPF | OPF |
|---|---|---|
| Whole operation cycle (5439 instants) | 389.5 | 21650.7 |
| Average value for a single instant | 0.07 | 3.98 |

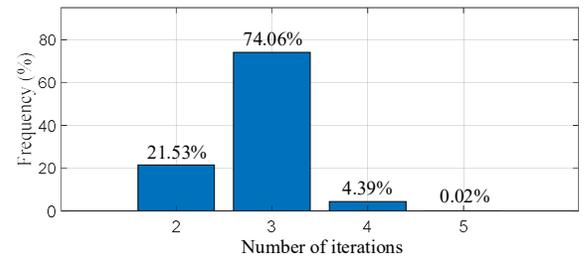

**Fig. 11.** Iteration times frequency statistics.

The speed up effect can be roughly interpreted as follows: Supposing there are $N$ TSSs and $M$ vehicles, the number of nodes is $(N+M)$. Usually, the calculation time of OPF is proportional to $(N+M)^3$. In the proposed algorithm, step 1 is a normal power flow calculation whose calculation time is proportional to $(N+M)^2$; the calculation time of steps 2, 3, and 4 is proportional to $N$. Hence, the calculation time of the proposed algorithm should be approximately $1/(N+M)$, i.e., 1/69, of the OPF's.

The frequency statistics for the number of iterations are shown in Fig. 11. Over 95% quasi-OPF calculation can be finished within 3 iterations, whereas OPF based on primal-dual interior point method needs approximately 50 iterations to converge. It is worth mentioning that the iterations of the proposed quasi-OPF are not due to repeated adjustment of optimization directions because quasi-OPF adopts a direct mapping from load information to near-optimal solutions. The iterations of the proposed quasi-OPF are due to the constant current sources modeling for rail vehicles in Section III: we assume rail vehicles nodes have constant current, as shown in Fig. 2, but actually rail vehicles nodes have constant power; therefore, the proposed algorithm needs two to five iterations to calculate the current value for current sources to ensure that the rail vehicle power is equal to the input value.

## VI. CONCLUSION

Flexible DC TPSs have many economic and technical merits based on state-of-the-art power electronics and system-level coordinated control. Fast OPF calculation is required to achieve better analysis, operations, and planning of flexible DC TPSs. In this study, first, we reveal the mechanism of OPF based on a new modeling method. The modeling method is based on the superposition principle, which decomposes an original system into the superposition of a coordinated control subsystem and a natural distribution subsystem. Second, a fast quasi-OPF algorithm is designed to solve near-optimal solutions based on the understanding of the internal mechanism of OPF. The effectiveness is verified by mathematical proofs and a case study with Beijing Metro Line 13. The algorithm simulates OPF with high computational efficiency, which achieves a speed-up of 57 times and solves near-optimal solutions in 0.07 s. Comparisons of the power flow in the whole operation cycle and the power flow in four typical instants are conducted to show the near-optimal performance of the quasi-OPF solutions.

## APPENDIX

*A. Parameters*

As a typical case, Beijing Metro No. 13 is a subway line in the reconstruction stage, after which its conventional TPS will be substituted by a flexible DC TPS, and the proposed quasi-OPF algorithm is promising for use in the control strategy design and planning of Beijing Metro No. 13.

The basic parameters are shown in TABLE III. There are 23 TSSs and 46 rail vehicles in the DC TPS. The station auxiliary loads of different TSSs vary, as shown in TABLE IV. It is apparent that the peak power demand of station auxiliary loads is much less than that of rail vehicles.

TABLE III
BEIJING METRO NO. 13 BASIC PARAMETERS

| Parameter | Value |
|---|---|
| Vehicle's maximum tractive power (MW) | 6.04 |
| Vehicle's maximum regenerative power (MW) | -9.58 |
| Rail vehicle maximum voltage (kV) | 0.95 |
| Rail vehicle maximum voltage without braking (kV) | 0.90 |
| Rail vehicle minimum voltage (kV) | 0.50 |
| VSC maximum allowable power (MW) | 11.00 |
| Efficiency of VSC | 98% |
| Resistance of third rail (ohm/km) | 0.0078 |
| Resistance of rail (ohm/km) | 0.02 |
| Minimum headway (min) | 2 |
| Dwell time (s) | 30-45 |

TABLE IV
STATION AUXILIARY LOADS OF TSSs

| TSS Sequence | Power (MW) | TSS Sequence | Power (MW) | TSS Sequence | Power (MW) |
|---|---|---|---|---|---|
| 1-3 | 0.54 | 13 | 0.10 | 21-22 | 0.54 |
| 4-10 | 0.21 | 14-18 | 0.54 | 23 | 0.66 |
| 11-12 | 0.35 | 19-20 | 0.10 | | |

*B. Proof of Proposition 1*

Use $dis$ to denote the location that varies from 0 to $L$. $I_c(dis)$, $I_{c\_nd}(dis)$, and $I_{c\_cc}(dis)$ denote the current in the catenary with respect to the location in the original system, natural distribution subsystem, and coordinated control subsystem, respectively. Based on the superposition principle, $I_c(dis)$ can be decomposed:

$$I_c(dis) = I_{c\_nd}(dis) + I_{c\_cc}(dis) \quad (21)$$

$I_{c\_nd}(dis)$ is the natural distribution of rail vehicles' current in the catenary, which is time-varying due to the change in rail vehicles' location or power demand. In the sense of probability statistics, $I_{c\_nd}(dis)$ can be considered a random variable; since the probability of the situation in which $I_{c\_nd}(dis)$ is positive (the current is toward one direction) and the probability of the situation in which $I_{c\_nd}(dis)$ is negative (the current is toward the other direction) is basically equal, the expectation $E(I_{c\_nd}(dis))$ is approximately 0, which has been further proven by simulation.

With the assumption that $E(I_{c\_nd}(dis)) = 0$, the expectation of power loss in the original system can be expressed as:

$$E(\int_0^L rI_c^2(dis)d(dis))$$
$$= E(\int_0^L r\left[I_{c\_nd}(dis) + I_{c\_cc}(dis)\right]^2 d(dis)) \quad (22)$$
$$= E(\int_0^L rI_{c\_nd}^2(dis)d(dis)) + \int_0^L rI_{c\_cc}^2(dis)d(dis)$$

Therefore, the expectation of power loss in the original system equals the summation of the expectation of power loss in the natural distribution subsystem and the expectation of power loss in the coordinated control subsystem. Because the expectation of power loss in the natural distribution subsystem is determined directly by loads, which is not controllable, optimizing power loss is equivalent to optimizing power loss in the coordinated control subsystem based on (22).

*C. Proof of Proposition 2*

Based on Kirchhoff's current law,

$$\sum_{i=1}^N I_{si\_cc} = 0 \quad (23)$$

Then, the following equations can be derived:

$$I_{s(m-1)\_cc} = -I_{sm\_cc} - I_{s(m+1)\_cc} - \sum_{i=1}^{m-2} I_{si\_cc} - \sum_{i=m+2}^{N} I_{si\_cc} \quad (24)$$

Based on Kirchhoff's current law, we know:
$$I_{c\_cc} = AI_{s\_cc} \quad (25)$$

where the matrix $A$ is defined by:
$$A = \begin{bmatrix} 1 & 0 & 0 & \cdots & 0 & 0 & 0 \\ 1 & 1 & 0 & \cdots & 0 & 0 & 0 \\ 1 & 1 & 1 & \cdots & 0 & 0 & 0 \\ \vdots & \vdots & \vdots & \ddots & \vdots & \vdots & \vdots \\ 1 & 1 & 1 & \cdots & 1 & 0 & 0 \\ 1 & 1 & 1 & \cdots & 1 & 1 & 0 \end{bmatrix} \quad (26)$$

Define the resistance matrix $R$:
$$R = diag(r_{c1}+r_{r1}, r_{c2}+r_{r2}, \ldots, r_{ci}+r_{ri}, \ldots, r_{c(N-1)}+r_{r(N-1)}) \quad (27)$$

For simplicity, we can use $r_i$ to represent $r_{ci}+r_{ri}$. The power loss in the coordinated control subsystem can be calculated by:
$$P_{loss} = I_{c\_cc}^T R I_{c\_cc} = I_{s\_cc}^T \widetilde{R} I_{s\_cc} \quad (28)$$

where
$$\widetilde{R} = A^T R A \quad (29)$$

$$\frac{dP_{loss}}{dI_{sm\_cc}} = -2 r_{m-1} \sum_{i=1}^{m-1} I_{si\_cc} \quad (30)$$

$$\frac{dP_{loss}}{dI_{s(m+1)\_cc}} = -2 \left[ (r_{m-1} + r_m) \sum_{i=1}^{m-1} I_{si\_cc} + r_m I_{sm\_cc} \right] \quad (31)$$

To optimize the power loss, the following equations should be satisfied:
$$\frac{dP_{loss}}{dI_{sm\_cc}} = 0, \quad \frac{dP_{loss}}{dI_{s(m+1)\_cc}} = 0 \quad (32)$$

Solving (32) based on (30) and (31), we can conclude
$$\sum_{i=1}^{m-1} I_{si\_cc} = 0, \quad \sum_{i=1}^{m} I_{si\_cc} = 0 \quad (33)$$

Based on (33), it is easy to conclude:
$$I_{sm\_cc} = 0 \quad (34)$$

*D. Proof of Corollary 1*

Based on (25), we can learn that
$$I_{rb(m-1)} = \sum_{i=1}^{m-1} I_{rci}, \quad I_{rbm} = \sum_{i=1}^{m} I_{rci} \quad (35)$$

Based on (33) and (35), we can conclude
$$I_{c(m-1)\_cc} = 0, \quad I_{cm\_cc} = 0 \quad (36)$$

*E. Proof of Proposition 3*

Assume the TSS $m$ to TSS $(m+z)$ needs coordinated support. Based on proposition 1, only TSS $(m-1)$ and $(m+z+1)$ provide coordinated support.

Based on Kirchhoff's current law, (23) still holds true. Then, the following equations can be derived:
$$I_{s(m-1)\_cc} = -I_{s(m+z+1)\_cc} - \sum_{i=1}^{m-2} I_{si\_cc} - \sum_{i=m}^{m+z} I_{si\_cc} - \sum_{i=m+z+2}^{N} I_{si\_cc} \quad (37)$$

$$\frac{dP_{loss}}{dI_{s(m+z+1)\_cc}} = -2 \left[ \sum_{i=m-1}^{m+z} r_i \sum_{i=1}^{m-1} I_{si\_cc} + \sum_{j=m}^{m+z} (I_{sj\_cc} \sum_{i=j}^{m+z} r_i) \right] \quad (38)$$

To optimize the power loss, the following equation should be satisfied:
$$\frac{dP_{loss}}{dI_{s(m+z+1)\_cc}} = 0 \quad (39)$$

Based on corollary 1, the first TSS to the $(m-2)$th TSS does not affect the coordinated support to the area from TSS $(m-1)$ to TSS $(m+z+1)$. Based on (38) and (39), we can derive that
$$\sum_{i=m-1}^{m+z} r_i I_{s(m-1)\_cc} + \sum_{j=m}^{m+z} (I_{sj\_cc} \sum_{i=j}^{m+z} r_i) = 0 \quad (40)$$

Then, we can conclude that
$$I_{s(m-1)\_cc} = -\left[ \sum_{j=m}^{m+z} (I_{sj\_cc} \sum_{i=j}^{m+z} r_i) \right] \bigg/ \sum_{i=m-1}^{m+z} r_i \quad (41)$$

which is the mathematical expression of Proposition 3.

REFERENCES

[1] J. Guo, S. Ma, T. Wang, Y. Jing, W. Hou and H. Xu, "Challenges of developing a power system with a high renewable energy proportion under China's carbon targets," iEnergy, vol. 1, no. 1, pp. 12-18, March 2022, doi: 10.23919/IEN.2022.0005.

[2] V. Gelman, "Insulated-Gate Bipolar Transistor Rectifiers: Why They Are Not Used in Traction Power Substations," IEEE Veh. Technol. Mag., vol. 9, no. 3, pp. 86–93, Sep. 2014, doi: 10.1109/MVT.2014.2333762.

[3] L. Hou et al., "A novel DC traction power supply system based on the modular multilevel converter suitable for energy feeding and de-icing," CSEE Journal of Power and Energy Systems, pp. 1–10, 2020, doi: 10.17775/CSEEJPES.2020.02260.

[4] E. Pilo de la Fuente, S. K. Mazumder, and I. G. Franco, "Railway Electrical Smart Grids: An introduction to next-generation railway power systems and their operation.," IEEE Electrification Magazine, vol. 2, no. 3, pp. 49–55, Sep. 2014, doi: 10.1109/MELE.2014.2338411.

[5] M. Brenna, F. Foiadelli, and H. J. Kaleybar, "The Evolution of Railway Power Supply Systems Toward Smart Microgrids: The concept of the energy hub and integration of distributed energy resources," IEEE Electrific. Mag., vol. 8, no. 1, pp. 12–23, Mar. 2020, doi: 10.1109/MELE.2019.2962886.

[6] A. Gomez-Exposito, J. M. Mauricio, and J. M. Maza-Ortega, "VSC-Based MVDC Railway Electrification System," IEEE Trans. Power Delivery, vol. 29, no. 1, pp. 422–431, Feb. 2014, doi: 10.1109/TPWRD.2013.2268692.

[7] ALSTOM. RE-USE Layman's Report September 2012–May 2018. Available online: https://www.alstom.com/sites/alstom.com/files/2018/10/30/re-use_laymans_report_en.pdf.

[8] A. Keane et al., "State-of-the-Art Techniques and Challenges Ahead for Distributed Generation Planning and Optimization," in IEEE Transactions on Power Systems, vol. 28, no. 2, pp. 1493-1502, May 2013, doi: 10.1109/TPWRS.2012.2214406.

[9] R. D. Zimmerman, C. E. Murillo-Sánchez and R. J. Thomas, "MATPOWER: Steady-State Operations, Planning, and Analysis Tools for Power Systems Research and Education," in IEEE Transactions on Power Systems, vol. 26, no. 1, pp. 12-19, Feb. 2011, doi: 10.1109/TPWRS.2010.2051168.

[10] Li, Hongbo, et al. "Energy conscious management for smart metro traction power supply system with 4G communication loop." Energy Reports 7 (2021): 798-807.

[11] G. Zhang, Z. Tian, P. Tricoli, S. Hillmansen, Y. Wang and Z. Liu, "Inverter Operating Characteristics Optimization for DC Traction Power Supply Systems," in IEEE Transactions on Vehicular Technology, vol. 68, no. 4, pp. 3400-3410, April 2019, doi: 10.1109/TVT.2019.2899165.

[12] V. A. Kleftakis and N. D. Hatziargyriou, "Optimal Control of Reversible Substations and Wayside Storage Devices for Voltage Stabilization and Energy Savings in Metro Railway Networks," in IEEE Transactions on Transportation Electrification, vol. 5, no. 2, pp. 515-523, June 2019, doi: 10.1109/TTE.2019.2913355.

[13] F. Hao, G. Zhang, J. Chen, Z. Liu, D. Xu and Y. Wang, "Optimal Voltage Regulation and Power Sharing in Traction Power Systems With Reversible Converters," in IEEE Transactions on Power Systems, vol. 35, no. 4, pp. 2726-2735, July 2020, doi: 10.1109/TPWRS.2020.2968108.

[14] F. Hao, G. Zhang, J. Chen and Z. Liu, "Distributed Reactive Power Compensation Method in DC Traction Power Systems With Reversible Substations," IEEE Transactions on Vehicular Technology, vol. 70, no. 10, pp. 9935-9944, Oct. 2021, doi: 10.1109/TVT.2021.3108030.

[15] Zhang, G., Tian, Z., Tricoli, P., Hillmansen, S., and Liu, Z.. "A new hybrid simulation integrating transient-state and steady-state models for the

analysis of reversible DC traction power systems." International Journal of Electrical Power & Energy Systems, vol. 109, pp. 9-19, July 2019.
[16] J. Zhu, X. Mo, Y. Xia, Y. Guo, J. Chen and M. Liu, "Fully-Decentralized Optimal Power Flow of Multi-Area Power Systems Based on Parallel Dual Dynamic Programming," in IEEE Transactions on Power Systems, vol. 37, no. 2, pp. 927-941, March 2022, doi: 10.1109/TPWRS.2021.3098812.
[17] A. Agarwal and L. Pileggi, "Large Scale Multi-Period Optimal Power Flow With Energy Storage Systems Using Differential Dynamic Programming," in IEEE Transactions on Power Systems, vol. 37, no. 3, pp. 1750-1759, May 2022, doi: 10.1109/TPWRS.2021.3115636.
[18] Z. Guo, W. Wei, L. Chen, Z. Dong and S. Mei, "Parametric Distribution Optimal Power Flow With Variable Renewable Generation," in IEEE Transactions on Power Systems, vol. 37, no. 3, pp. 1831-1841, May 2022, doi: 10.1109/TPWRS.2021.3110528.
[19] Z. Yan and Y. Xu, "A Hybrid Data-driven Method for Fast Solution of Security-Constrained Optimal Power Flow," in IEEE Transactions on Power Systems, doi: 10.1109/TPWRS.2022.3150023.
[20] W. Huang, X. Pan, M. Chen and S. H. Low, "DeepOPF-V: Solving AC-OPF Problems Efficiently," in IEEE Transactions on Power Systems, vol. 37, no. 1, pp. 800-803, Jan. 2022, doi: 10.1109/TPWRS.2021.3114092.
[21] T. Falconer and L. Mones, "Leveraging power grid topology in machine learning assisted optimal power flow," in IEEE Transactions on Power Systems, 2022, doi: 10.1109/TPWRS.2022.3187218.
[22] Y. Chen, S. Lakshminarayana, C. Maple and H. V. Poor, "A Meta-Learning Approach to the Optimal Power Flow Problem Under Topology Reconfigurations," in IEEE Open Access Journal of Power and Energy, vol. 9, pp. 109-120, 2022, doi: 10.1109/OAJPE.2022.3140314.
[23] L. Gan and S. H. Low, "Optimal Power Flow in Direct Current Networks," in IEEE Transactions on Power Systems, vol. 29, no. 6, pp. 2892-2904, Nov. 2014, doi: 10.1109/TPWRS.2014.2313514.
[24] H. M. A. Ahmed and M. M. A. Salama, "Energy Management of AC–DC Hybrid Distribution Systems Considering Network Reconfiguration," in IEEE Transactions on Power Systems, vol. 34, no. 6, pp. 4583-4594, Nov. 2019, doi: 10.1109/TPWRS.2019.2916227.
[25] Z. Yang, H. Zhong, A. Bose, Q. Xia and C. Kang, "Optimal Power Flow in AC–DC Grids With Discrete Control Devices," in IEEE Transactions on Power Systems, vol. 33, no. 2, pp. 1461-1472, March 2018, doi: 10.1109/TPWRS.2017.2721971.
[26] H. -T. Zhang, W. Sun, Y. Li, D. Fu and Y. Yuan, "A Fast Optimal Power Flow Algorithm Using Powerball Method," in IEEE Transactions on Industrial Informatics, vol. 16, no. 11, pp. 6993-7003, Nov. 2020, doi: 10.1109/TII.2019.2909328.
[27] Y. Tang, K. Dvijotham and S. Low, "Real-Time Optimal Power Flow," in IEEE Transactions on Smart Grid, vol. 8, no. 6, pp. 2963-2973, Nov. 2017, doi: 10.1109/TSG.2017.2704922.
[28] E. Dall'Anese and A. Simonetto, "Optimal Power Flow Pursuit," in IEEE Transactions on Smart Grid, vol. 9, no. 2, pp. 942-952, March 2018, doi: 10.1109/TSG.2016.2571982.
[29] A. Bernstein and E. Dall'Anese, "Real-Time Feedback-Based Optimization of Distribution Grids: A Unified Approach," in IEEE Transactions on Control of Network Systems, vol. 6, no. 3, pp. 1197-1209, Sept. 2019, doi: 10.1109/TCNS.2019.2929648.
[30] Y. Yao, F. Ding, K. Horowitz and A. Jain, "Coordinated Inverter Control to Increase Dynamic PV Hosting Capacity: A Real-Time Optimal Power Flow Approach," in IEEE Systems Journal, vol. 16, no. 2, pp. 1933-1944, June 2022, doi: 10.1109/JSYST.2021.3071998.
[31] A. O. Rousis, I. Konstantelos and G. Strbac, "A Planning Model for a Hybrid AC–DC Microgrid Using a Novel GA/AC OPF Algorithm," in IEEE Transactions on Power Systems, vol. 35, no. 1, pp. 227-237, Jan. 2020, doi: 10.1109/TPWRS.2019.2924137.
[32] G. Muñoz-Delgado, J. Contreras and J. M. Arroyo, "Distribution System Expansion Planning Considering Non-Utility-Owned DG and an Independent Distribution System Operator," in IEEE Transactions on Power Systems, vol. 34, no. 4, pp. 2588-2597, July 2019, doi: 10.1109/TPWRS.2019.2897869.
[33] X. Xu, J. Li, Y. Xu, Z. Xu and C. S. Lai, "A Two-Stage Game-Theoretic Method for Residential PV Panels Planning Considering Energy Sharing Mechanism," in IEEE Transactions on Power Systems, vol. 35, no. 5, pp. 3562-3573, Sept. 2020, doi: 10.1109/TPWRS.2020.2985765.
[34] M. Kabirifar, M. Fotuhi-Firuzabad, M. Moeini-Aghtaie, N. Pourghaderi and P. Dehghanian, "A Bi-Level Framework for Expansion Planning in Active Power Distribution Networks," in IEEE Transactions on Power Systems, vol. 37, no. 4, pp. 2639-2654, July 2022, doi: 10.1109/TPWRS.2021.3130339.
[35] P. Arboleya, B. Mohamed and I. El-Sayed, "DC Railway Simulation Including Controllable Power Electronic and Energy Storage Devices," in IEEE Transactions on Power Systems, vol. 33, no. 5, pp. 5319-5329, Sept. 2018, doi: 10.1109/TPWRS.2018.2801023.
[36] R. A. Jabr and I. Džafić, "Solution of DC Railway Traction Power Flow Systems Including Limited Network Receptivity," in IEEE Transactions on Power Systems, vol. 33, no. 1, pp. 962-969, Jan. 2018, doi: 10.1109/TPWRS.2017.2688338.